\documentclass[12pt,aps,preprint,nofootinbib]{revtex4}

\usepackage{epsfig}
\usepackage{amssymb}
\usepackage{amsmath}
\usepackage{amsfonts}
\usepackage{graphics}
\usepackage{cancel}
\usepackage{bbm}

\newcommand{\eps}{\varepsilon}
\newcommand{\Refs}{Refs.}
\newcommand{\Ref}{Ref.}
\newcommand{\Sec}{Sec.}

\newcommand{\Tab}{Tab.}
\newcommand{\App}{App.}
\newcommand{\eq}{Eq.}

\newcommand{\Fig}{Fig.}

\newcommand{\bea}{\begin{eqnarray}}
\newcommand{\eea}{\end{eqnarray}}

\newcommand{\be}{\begin{equation}}
\newcommand{\ee}{\end{equation}}

\newcommand{\ba}{\begin{array}}
\newcommand{\ea}{\end{array}}

\newcommand{\ie}{\emph{i.e.}}
\newcommand{\eg}{\emph{e.g.}}

\newcommand{\CP}{{CP}}
\newcommand{\hc}{\mathrm{H.c.}}

\DeclareMathOperator{\diag}{diag}
\DeclareMathOperator{\Real}{Re}
\DeclareMathOperator{\Imag}{Im}

\allowdisplaybreaks

\begin{document}

\title{Probing non-unitary mixing and CP-violation at a\\Neutrino Factory}

\author{Stefan Antusch}
\email[]{antusch@mppmu.mpg.de}
\affiliation{Max-Planck-Institut f\"ur Physik (Werner-Heisenberg-Institut),
F\"ohringer Ring 6, 80805 M\"unchen, Germany}
\author{Mattias Blennow}
\email[]{blennow@mppmu.mpg.de}
\affiliation{Max-Planck-Institut f\"ur Physik (Werner-Heisenberg-Institut),
F\"ohringer Ring 6, 80805 M\"unchen, Germany}
\author{Enrique Fernandez-Martinez}
\email[]{enfmarti@mppmu.mpg.de}
\affiliation{Max-Planck-Institut f\"ur Physik (Werner-Heisenberg-Institut),
F\"ohringer Ring 6, 80805 M\"unchen, Germany}
\author{Jacobo L\'opez-Pav\'on}
\email[]{jacobo.lopez@uam.es}
\affiliation{Instituto de F\'\i sica Te\'orica UAM/CSIC, 
Universidad Aut\'onoma de Madrid, 28049 Cantoblanco, Madrid, Spain}

\begin{abstract}
A low energy non-unitary leptonic mixing matrix is a generic feature
of many extensions of the Standard Model. In such a case, the task of
future precision neutrino oscillation experiments is more ambitious
than measuring the three mixing angles and the leptonic (Dirac)
\CP-phase, \ie, the accessible parameters of a unitary leptonic mixing
matrix. A non-unitary mixing matrix has 13 parameters that affect
neutrino oscillations, out of which four are \CP-violating. In the
scheme of Minimal Unitarity Violation (MUV) we analyse the potential
of a Neutrino Factory for determining or constraining the parameters
of the non-unitary leptonic mixing matrix, thereby testing the origin
of \CP-violation in the lepton sector.
\end{abstract}

\pacs{}

\preprint{MPP-2009-31}
\preprint{IFT-UAM/CSIC-09-16}
\preprint{EURONU-WP6-09-04}

\maketitle

\section{Introduction}
There are several indications from particle physics, as well as from
cosmology, for the existence of physics beyond the Standard Model
(SM). For example, the gauge hierarchy problem suggests that new
physics exists at energies close to the electroweak scale in order to
stabilise it against large quantum corrections. In cosmology, the
evidence for dark matter in the Universe requires the extension of the
SM particle content. Last, but not least, the discovery that neutrinos
are massive provides the first clear particle physics evidence that
the SM has to be extended.

In general, extensions of the SM will also affect the physics relevant
at neutrino oscillation experiments. New physics effects on neutrino
oscillations are particularly relevant for the next generation of
precision neutrino oscillation facilities such as Neutrino
Factories~\cite{Geer:1997iz,DeRujula:1998hd}, which aim at measuring
the unknown leptonic mixing angle $\theta_{13}$, the neutrino mass
hierarchy (\ie, $\mbox{sgn}(\Delta m_{31}^2)$), as well as the Dirac
phase $\delta$, which can induce \CP-violation in neutrino
oscillations. In most phenomenological studies regarding the
sensitivities of future neutrino oscillation facilities, the leptonic
mixing matrix is assumed to be unitary.

In contrast to this common practice, it is well known that one generic
feature of new physics in the lepton sector is the non-unitarity of
the low energy leptonic mixing matrix. This non-unitarity appears
whenever additional heavy particles mix with the light neutrinos or
their charged lepton partners \cite{Langacker:1988ur}. After
integrating the heavy states out of the theory, the $3 \times 3$
submatrix of the light neutrinos remains as an effective mixing
matrix. This low energy leptonic mixing matrix is, in general, not
unitary.

While there are many models of physics beyond the SM which induce
non-unitarity, an extension of the SM featuring a non-unitary leptonic
mixing can be described in a minimal way through an effective theory,
the so-called \emph{Minimal Unitarity Violation (MUV)}
scheme~\cite{Antusch:2006vwa}. It contains the relevant low-energy
information for neutrino oscillation experiments and is minimal in the
sense that only three light neutrinos are considered and that new
physics is introduced in the neutrino sector only. It provides an
effective description of all models where additional heavy singlets
mix with three light neutrinos.\footnote{Other possibilities to
  introduce non-unitary leptonic mixing are, \eg, via an additional
  vector-like lepton generation or via fermionic SU(2)$_\mathrm{L}$
  triplets, which are beyond MUV. Non-unitarity in these schemes turns
  out to be significantly more constrained by non-oscillation
  experiments than in MUV (see, \eg, \Ref~\cite{Abada:2007ux}).}

In MUV, the charged- and neutral-current interactions of the neutrinos
(\ie, their couplings to the $W$ and $Z$ bosons) are modified. The
non-unitary leptonic mixing matrix $N$, which appears in the charged-current
interaction, contains the only additional degrees of freedom, since the
neutral-current interaction of the neutrinos is proportional to
$N^\dagger N$ while the neutral-current interaction of the charged
leptons is unchanged. Thus, instead of the three mixing angles and
three \CP-phases of a unitary leptonic mixing matrix (with only one
affecting neutrino oscillations), the non-unitary mixing matrix $N$
contains 15 parameters, out of which six are \CP-violating phases
(including two Majorana phases, which do not affect neutrino
oscillations).
 
In this study, we investigate the potential of a Neutrino Factory for
determining or constraining the parameters of the non-unitary leptonic
mixing matrix, thereby testing the origin of \CP-violation in the
lepton sector.

\section{General introduction to unitarity violation}
\label{sec:generalUV}

As motivated in the introduction, non-unitarity of the leptonic mixing
matrix is a generic manifestation of new physics in the lepton
sector. The MUV scheme provides an effective field theory extension of
the SM and is minimal in the sense that only three light neutrinos are
considered and that new physics is only introduced in the neutrino
sector. Notice that this assumption is conservative, since new physics
affecting other sectors, such as that of the charged leptons, will
lead to stronger signals than the ones discussed here. The MUV scheme
thus describes the relevant effects on neutrino oscillations in the
various types of models where the SM is extended by heavy
singlet fermions (where ``heavy'' refers to large masses compared to
the energies of the neutrino oscillation experiments) which mix with
the light neutrinos.

In the MUV scheme, the Lagrange density of the SM is extended by two
effective operators, one of mass dimension five and one of mass
dimension six. The dimension five operator is the ubiquitous lepton
number violating Weinberg operator $\delta{\cal L}^{d=5} =
\frac{1}{2}\, c_{\alpha \beta}^{d=5} \,\left( \overline{L^c}_{\alpha}
\tilde \phi^* \right) \left( \tilde \phi^\dagger \, L_{ \beta} \right)
+ \hc$, the lowest dimensional effective operator for generating
neutrino masses using the field content of the SM. The coefficient
matrix $c_{\alpha \beta}^{d=5}$ is of $\mathcal{O}(1/M)$ and related
to the low energy neutrino mass matrix by $m_\nu = v^2_{\mathrm{EW}}
c_{}^{d=5}$, where $v_{\mathrm{EW}}$ is the vacuum expectation value
of the SM Higgs field $\phi$, which breaks the electroweak symmetry,
and $\tilde \phi = i \tau_2 \phi^*$. The SM neutrinos are contained in
the lepton doublets $L_\alpha$, with $\alpha = e,\mu,\tau$ running
over the three families.

The effective dimension six operator $c^{d=6}_{\alpha \beta} \, \left(
\overline{L}_{\alpha} \tilde \phi \right) i \cancel{\partial}
\left(\tilde \phi^\dagger L_{ \beta} \right)$ conserves lepton
number\footnote{We note that since the dimension six operator
  conserves lepton number, it is not necessarily suppressed by the
  smallness of the neutrino masses.} and, after electroweak symmetry
breaking, contributes to the kinetic terms of the neutrinos. After
their canonical normalisation, they generate a non-unitary leptonic
mixing matrix $N$, as well as non-universal couplings of the neutrinos
to the $Z$ boson proportional to $N^\dagger N$. The modified part of
the Lagrange density in MUV is given by
\begin{eqnarray}
\label{eff-lagr}
{\cal L}^{\rm eff}
&=&
\frac{1}{2}\,\left(
\bar{\nu}_ii\,{\partial\hspace{-6pt}\slash}\,\nu_i
-\,
\overline{{\nu}^c}_im_{i}\,\nu_i  +\,\hc \right)\,
-\,\frac{g}{2\sqrt{2}}\,
 ( W^+_\mu\,\bar{l}_\alpha\,\gamma_\mu\,(1-\gamma_5)\,N_{\alpha i}\,\nu_i
   \,+\hc) \nonumber \\
&&-\,\frac{g}{2 \cos\theta_W}\,
 ( Z_\mu\,\bar{\nu}_i\,\gamma^\mu\,(1-\gamma_5)\,
     (N^{\dagger}N)_{ij}\,\nu_j\,+\,\hc)\,
 \,. 
\end{eqnarray}
We note that the MUV scheme is also minimal in the sense that all new
physics effects depend on the non-unitary leptonic mixing matrix
$N$. Regarding neutrino oscillation experiments, the non-unitarity of
$N$ affects the processes at the source and the detector as well as
neutrino propagation in matter.

To parametrise $N$, we use the fact that a general matrix can be
written as the product of a Hermitian matrix times a unitary
matrix. Decomposing the Hermitian matrix as $\mathbbm{1} + \eps$ (with
$\eps = \eps^\dagger$) and denoting the unitary matrix by $U$, we can
write \cite{FernandezMartinez:2007ms}
\begin{equation}
N = (\mathbbm{1} + \eps)\,U\;.
\label{param}
\end{equation}
For the complex off-diagonal elements of the matrix $\eps$, we use the
notation $\eps_{\alpha\beta} = |\eps_{\alpha\beta}| e^{i
  \phi_{\alpha\beta}}$. Notice that, due to the Hermiticity of $\eps$,
$|\eps_{\alpha\beta}| = |\eps_{\beta\alpha}|$ and $\phi_{\alpha\beta}
= - \phi_{\beta\alpha}$. The diagonal elements are real and no further
parametrisation is required. Constraints on the $\eps_{\alpha\beta}$
can also be derived from the experimental data on electroweak decays
\cite{Nardi:1994iv,Tommasini:1995ii}. The present 90~\% CL bounds are
$|\eps_{\mu e}| < 3.5 \cdot 10^{-5}$, $|\eps_{\tau e}| < 8.0 \cdot
10^{-3}$, $|\eps_{\tau \mu}| < 5.1 \cdot
10^{-3}$~\cite{Antusch:2006vwa} and $|\eps_{ee}| < 2.0 \cdot 10^{-3}$,
$|\eps_{\mu\mu}| < 8.0 \cdot 10^{-4}$, $|\eps_{\tau\tau}| < 2.7 \cdot
10^{-3}$~\cite{Antusch:2008tz}. In our analysis, we will consider
unitarity violation consistent with the present bounds. Analytic
expressions for the neutrino oscillation probabilities in terms of $U$
and $\eps$ can be found in \App~\ref{sec:appendix}.

Finally, we would like to comment on other possible parametrisations
of a non-unitary leptonic mixing matrix. In
\Refs~\cite{Bekman:2002zk,Holeczek:2007kk,Xing:2007zj,Xing:2009vb}, a
different parametrisation is advocated, in which the deviations from
unitarity of the mixing matrix involving the three light neutrinos is
related to the mixing between these light neutrinos and the heavy
singlets in seesaw type theories.  The mixing matrix in a seesaw
scenario is the unitary matrix that diagonalises the extended neutrino
mass matrix:
\begin{equation} 
U_{6 \times 6}^T \left(
\begin{array}{cc}
0 & m_D \\
m_D^T & M_N
\end{array}
\right)
U_{6 \times 6} =
\left( 
\begin{array}{cc}
m & 0 \\
0 & M
\end{array}
\right),
\label{diag}
\end{equation}
where $m_D$ and $M_N$ are the neutrino's Dirac and Majorana mass
matrices, respectively. In the case of only one neutrino family, the
unitary matrix is just a rotation of angle $\theta \simeq m_D/M$. The
extension to three or more families is straightforward, performing the
diagonalisation in two steps: first a block-diagonalisation and then
two unitary rotations to diagonalize the mass matrices of the light and
heavy neutrinos, \ie,
\begin{equation}
U_{6 \times 6} = \left(
\begin{array}{cc}
A & B \\
C & D
\end{array}
\right)
\left( 
\begin{array}{cc}
U & 0 \\
0 & V
\end{array}
\right),
\label{block}
\end{equation}
where $U$ and $V$ are unitary matrices. Without loss of generality, we
can choose a basis for the heavy singlets such that $V=I$. Analogously
to the one family example, when performing the block diagonalisation,
the mixing between the light and heavy neutrinos is suppressed so that
\begin{equation}
 B \simeq \Theta = m_D M_N^{-1}.
 \label{theta}
\end{equation} 
This suppression is exploited in \Refs~\cite{Xing:2007zj,Xing:2009vb},
where the block diagonalising matrix is written as the product of the
9 possible rotations mixing the light and heavy states and then
expanded up to second order in the small mixing angles. This results
in
\begin{eqnarray}
A 
&=&
1 - \left( 
\begin{array}{ccc}
 \frac{1}{2} \left( s^2_{14} +
s^2_{15} + s^2_{16} \right) & 0 & 0 \\ \hat{s}^{}_{14}
\hat{s}^*_{24} + \hat{s}^{}_{15} \hat{s}^*_{25} + \hat{s}^{}_{16}
\hat{s}^*_{26} & \frac{1}{2} \left( s^2_{24} + s^2_{25} + s^2_{26}
\right) & 0 \\ \hat{s}^{}_{14} \hat{s}^*_{34} + \hat{s}^{}_{15}
\hat{s}^*_{35} + \hat{s}^{}_{16} \hat{s}^*_{36} & \hat{s}^{}_{24}
\hat{s}^*_{34} + \hat{s}^{}_{25} \hat{s}^*_{35} + \hat{s}^{}_{26}
\hat{s}^*_{36} & \frac{1}{2} \left( s^2_{34} +
s^2_{35} + s^2_{36} \right)
\end{array}
\right) + {\cal O}(\theta^4_{ij}) \; , \nonumber \\
B 
&=&
\left( 
\begin{array}{ccc} 
\hat{s}^*_{14} & \hat{s}^*_{15}
& \hat{s}^*_{16} \\ \hat{s}^*_{24} & \hat{s}^*_{25} &
\hat{s}^*_{26} \\ \hat{s}^*_{34} & \hat{s}^*_{35} &
\hat{s}^*_{36} 
\end{array}
\right) + {\cal O}(\theta^3_{ij}) \; ,
\label{xing}
\end{eqnarray}
where $\hat s_{ij} = s_{ij}\exp(i\delta_{ij})$ and $s_{ij} =
\sin(\theta_{ij})$. Notice that the mixing matrix of the three light
neutrinos is given by $N=A U$. Thus, the deviation from unitarity,
encoded in $A$, is directly related to the mixing $B$ between the
heavy and light neutrinos. We argue that this is also the case with
the Hermitian unitarity deviation adopted in
\eq~(\ref{param}). Indeed, we can exploit the suppression of
\eq~(\ref{theta}) to write the unitary block diagonalisation as the
exponential expansion of an anti-Hermitian matrix:
\begin{equation}
\left(
\begin{array}{cc}
A & B \\
C & D
\end{array}
\right) =
\exp
\left( 
\begin{array}{cc}
0 & \Theta \\
-\Theta^\dagger & 0
\end{array}
\right) =
\left( 
\begin{array}{cc}
1-\frac{1}{2}\Theta \Theta^\dagger & \Theta \\
-\Theta^\dagger & 1-\frac{1}{2}\Theta^\dagger \Theta
\end{array}
\right) + {\cal O}(\Theta^3).  
\end{equation} 
Thus, the Hermitian deviation from unitarity defined in
\eq~(\ref{param}) is just $\eps = -\Theta \Theta^\dagger/2$ and its
relation to the mixing between light and heavy neutrinos in a seesaw
scenario is straightforward.\footnote{The anti-Hermitian part can be
  reabsorbed in the unitary rotation, and is thus related to using
  different parametrisations.} Furthermore, notice that the deviation
from unitary mixing parametrised as in \eq~(\ref{xing}) can only be
applied to the specific case of the mixing between three light and
three heavy neutrinos while the product of an Hermitian and a unitary
matrix is a completely general matrix and thus suitable to take into
account more general scenarios. In addition, the unitarity deviation
$\eps$ is given by the coefficient of the $d=6$ operator ($\eps =
-c_{}^{d=6}/2$) that modifies the neutrino kinetic terms, introduced
in the MUV scheme and obtained in the effective theory of the seesaw
mechanism after integrating out the heavy singlets (see, \eg,
\Ref~\cite{Broncano:2002rw}).

\section{Numerical simulation and results}

We will now discuss the sensitivity of future neutrino oscillation
experiments to the different parameters of the MUV scheme. In
particular, we study the Neutrino Factory setup proposed in the
International Design Study (IDS) \cite{Bandyopadhyay:2007kx,IDS},
which consists of $\nu_e$ and $\nu_\mu$ beams from $5\cdot 10^{20}$
muon decays per year per baseline. We consider a setting where the
experiment is assumed to run for five years in each polarity. The
parent muons are assumed to have an energy of 25~GeV. The beams are
detected at two far sites, the first located at 4000~km with a 50~kton
Magnetised Iron Neutrino Detector (MIND) \cite{Abe:2007bi} and a
10~kton Emulsion Cloud Chamber (ECC) for $\tau$ detection
\cite{Donini:2002rm,Autiero:2003fu}, and the second located close to
the magic baseline~\cite{BurguetCastell:2001ez,Huber:2003ak} at
7500~km with an iron detector identical to the one at 4000~km.

A clean signal of a non-unitary mixing is the presence of
``zero-distance effects'' stemming from the non-orthogonality of the
flavour states.  Indeed, if the flavour basis is not orthogonal, a
neutrino of flavour $\alpha$ can be detected with flavour $\beta$
without the need of flavour conversion in the propagation.  This
translates to a baseline-independent term in the oscillation
probabilities, which is best probed at short distances, since the flux
is larger and it cannot be hidden by the standard oscillations. For
short baselines, this term is ($\alpha \neq \beta$)
\begin{equation} P_{\alpha
  \beta}(L=0) = 4 |\eps_{\alpha \beta}|^2 + \mathcal{O}(\eps^3).
\end{equation}
The oscillation probabilities for longer baselines up to second order
in the small parameters are derived in \App~\ref{sec:appendix}. Near
detectors are thus excellent for probing the zero-distance effect, in
particular $\tau$ detectors are of importance, since the present
bounds on $\eps_{\mu e}$ and $\eps_{\mu \mu}$ are rather strong. We
will therefore study the impact of near $\tau$ detectors of different
sizes located at 1~km from the beam source. In particular, we will
present all the results for near detector sizes of 100~ton, 1~kton,
and 10~kton, as well as the results without any near $\tau$
detector. Notice that 10~kton is the detector mass discussed for the
ECC detector located at 4000~km. However, we have seen no improvement
adding such a detector at that baseline while the gain in sensitivity
that a near detector capable of $\tau$ detection can provide is
significant, as we will discuss below. Therefore, we also considered
the larger mass to show what could be achieved with the planned
10~kton detector located at 1~km instead of 4000~km. To simulate the
near detector, we use the point-source and far-distance
approximations. These assumptions are reasonable, although somewhat
optimistic in the high-energy region, as can be seen in \Fig~12 of
\Ref~\cite{Tang:2009na}. However, the loss of flux at higher energies,
which corresponds to the on-axis neutrinos, may be recovered by using
rather elongated geometries of the near detector.  These are precisely
the kind of geometries that are being discussed for a magnetized
version of the ECC (MECC). Such a detector would be limited in size by
the above mentioned geometrical considerations and is not likely to be
larger than 4~kton. On the other hand, all the decay channels of the
$\tau$ could be studied in the magnetized version, which would
translate into an increase of the efficiency by a factor 5 with
respect to the ECC search for $\tau$ decays into $\mu$ considered
here. The impact of near $\mu$ detectors is still essentially to
normalise the neutrino flux and cross-sections, since the bounds on
$\eps_{\mu\mu}$ and $\eps_{\mu e}$ from the unitarity of the CKM
matrix and $\mu \to e \gamma$ are particularly strong
\cite{Antusch:2006vwa,Antusch:2008tz}.

In our simulations, we will study the ``golden'' \cite{Cervera:2000kp}
$\nu_e \to \nu_\mu$ and $\nu_\mu$ disappearance channels in the MIND
detectors and the ``silver'' \cite{Donini:2002rm,Autiero:2003fu}
$\nu_e \to \nu_\tau$ and ``discovery'' \cite{Donini:2008wz} $\nu_\mu
\to \nu_\tau$ channels at the ECC detectors, both near and far. For
the detector efficiencies and backgrounds, we follow the study in
\Ref~\cite{Abe:2007bi} of the MIND detector exposed to the Neutrino
Factory beam. The efficiencies and backgrounds for the silver channel
with an ECC detector are carefully discussed in
\Ref~\cite{Autiero:2003fu} and we follow the results of that
reference. Lacking an analogous study for the discovery channel, we
assume the same efficiencies and backgrounds as the ones for the
silver channel described in \Ref~\cite{Autiero:2003fu}.

For our numerical simulations, we scan the complete MUV parameter
space, adding nine unitarity violating parameters to the six standard
neutrino oscillation parameters. The scan is performed using the
MonteCUBES software \cite{mcubes,mcubeshome}, which allows to perform
Markov Chain Monte Carlo (MCMC) simulations with GLoBES
\cite{Huber:2004ka,Huber:2007ji}. For the implementation of the
unitarity deviations in the neutrino oscillation probabilities, we use
the NonUnitarity Engine (NUE) distributed along with the MonteCUBES
package. Using the MCMC technique allows the study of possible
parameter correlations in the full parameter space without restricting
the search to varying only a small subset of the parameters. This is
due to the fact that the number of evaluations required by Monte Carlo
techniques increases at most polynomially with the number of
parameters, while a scan based on grids in the parameter space would
require to evaluate the event rates and likelihoods at a number of
points that grows exponentially. For all of our figures, we have used
simulations with four MCMC chains containing $2 \times 10^6$ samples
each. In addition, we have checked that the chains have reached proper
convergence, in all cases better than $R-1 = 10^{-2}$
\cite{Gelman:1992}. It is also important to note that, unlike in the
standard usage of the GLoBES software, the use of MCMC techniques is
based on Bayesian rather than frequentist parameter estimation and, as
such, the result depends on the adopted priors. As priors, we will
consider the current bounds on both the standard and the unitarity
violating parameters, except for parameters to which the Neutrino
Factory has superior sensitivity, for which we use flat priors.

Before discussing the more detailed studies, let us comment on some of
the general results from the simulations. First of all, one of the
most remarkable features is that the results do not contain
significant correlations between any of the unitarity violating
parameters, nor are the unitarity violating parameters significantly
correlated with the standard neutrino oscillation parameters. The only
exception are some mild correlations between $\theta_{13}$, $\delta$
and the modulus and phase of $\eps_{\tau e}$ in the absence of
near $\tau$ detectors which, however, do not lead to new degeneracies
between these parameters or spoil the determination of $\theta_{13}$
and $\delta$ at the Neutrino Factory. Furthermore, the addition of a
near $\tau$ detector of only 100~ton is enough to almost completely
erase these correlations.  This implies that the Neutrino Factory
setup considered here has enough sensitivity to distinguish the
effects induced by unitarity violation from changes in the standard
parameters. Second, the sensitivities of the Neutrino Factory to the
diagonal parameters of the $\eps$ matrix, as well as to $\eps_{\mu
  e}$, do not improve with respect to the bounds derived from
electroweak decays, which are too stringent to allow for observable
effects at the Neutrino Factory.  Notice that none of the oscillation
probabilities studied here depend on $\eps_{ee}$, as shown in
\App~\ref{sec:appendix}.

We will thus concentrate on the sensitivities to $\eps_{\tau\mu}$ and
$\eps_{\tau e}$ in the next subsections, even though the other
unitarity violating parameters and standard oscillation parameters are
allowed to vary in the simulations.  As an example of the
sensitivities and correlations to all the 15 parameters considered,
the 105 projections to the different two-dimensional subspaces and the
marginalized regions for the 15 parameters can be studied in a
triangle plot at \Ref~\cite{mcubeshome} for the case of no near $\tau$
detector. The input values chosen for the unknown parameters in this
example were $\theta_{13}=5^\circ$, $\delta=0$, $|\eps_{\tau e}| =
0.005$ and $\phi_{\tau e} = \pi/4$, the input for the rest of the
non-unitary parameters was set to zero.  In all our simulations we
assume \cite{Maltoni:2004ei,GonzalezGarcia:2007ib}
$\theta_{12}=33^\circ$, $\theta_{23}=45^\circ$, $\Delta m^2_{21}= 8
\cdot 10^{-5}$~eV$^2$ and $\Delta m^2_{31}= 2.6 \cdot 10^{-3}$~eV$^2$.
We also assumed $4$~\% priors on $\theta_{12}$ and $\Delta m^2_{21}$
at $1 \sigma$, flat priors were used for the rest of the standard
oscillation parameters. For the unitarity violating parameters, we
consider Gaussian priors given by the ranges mentioned in
\Sec~\ref{sec:generalUV}.

\subsection{Sensitivity to $\boldsymbol{\eps_{\tau\mu}}$}

\begin{figure}
 \begin{center}
  \includegraphics[width=0.49\textwidth]{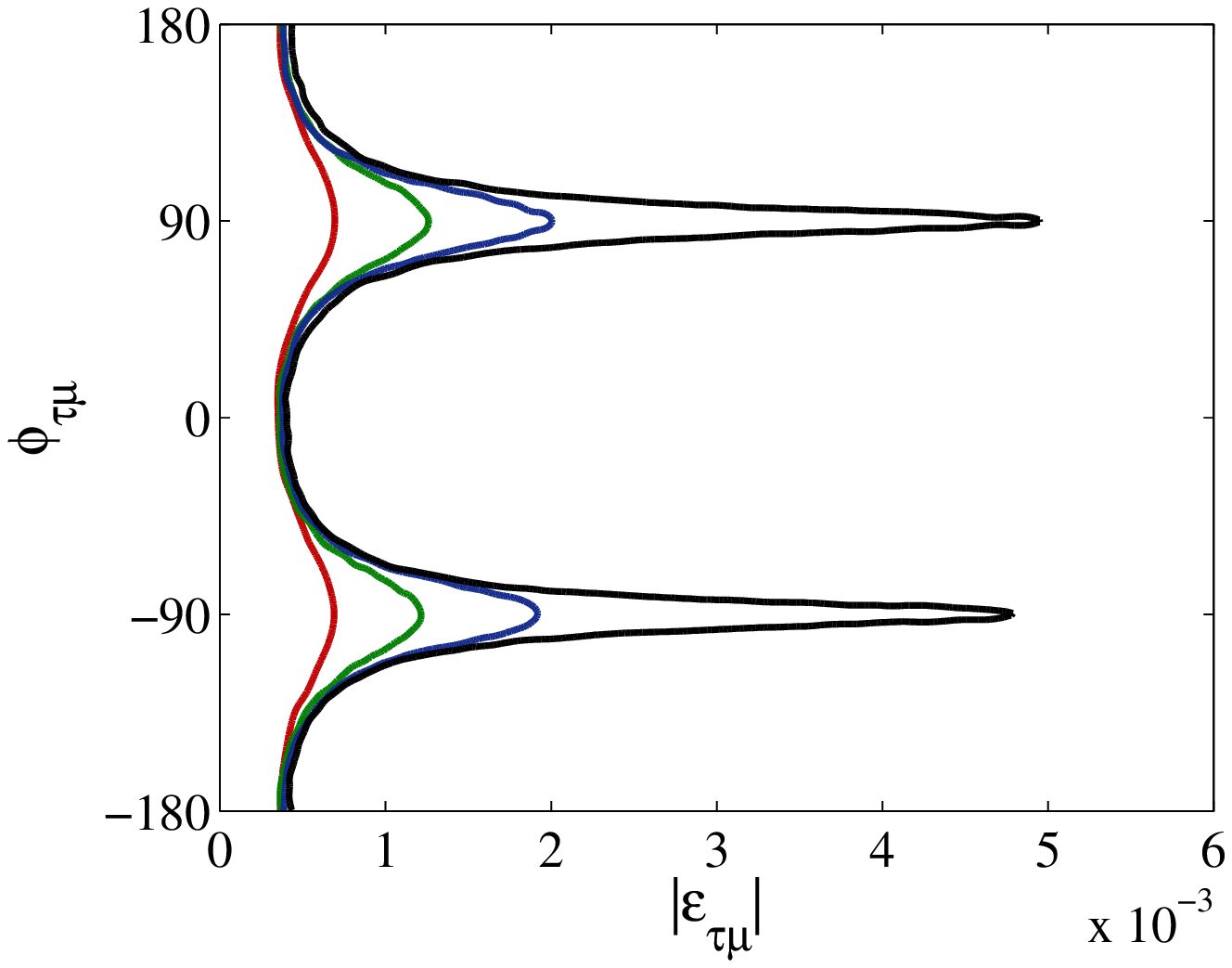}
  \includegraphics[width=0.49\textwidth]{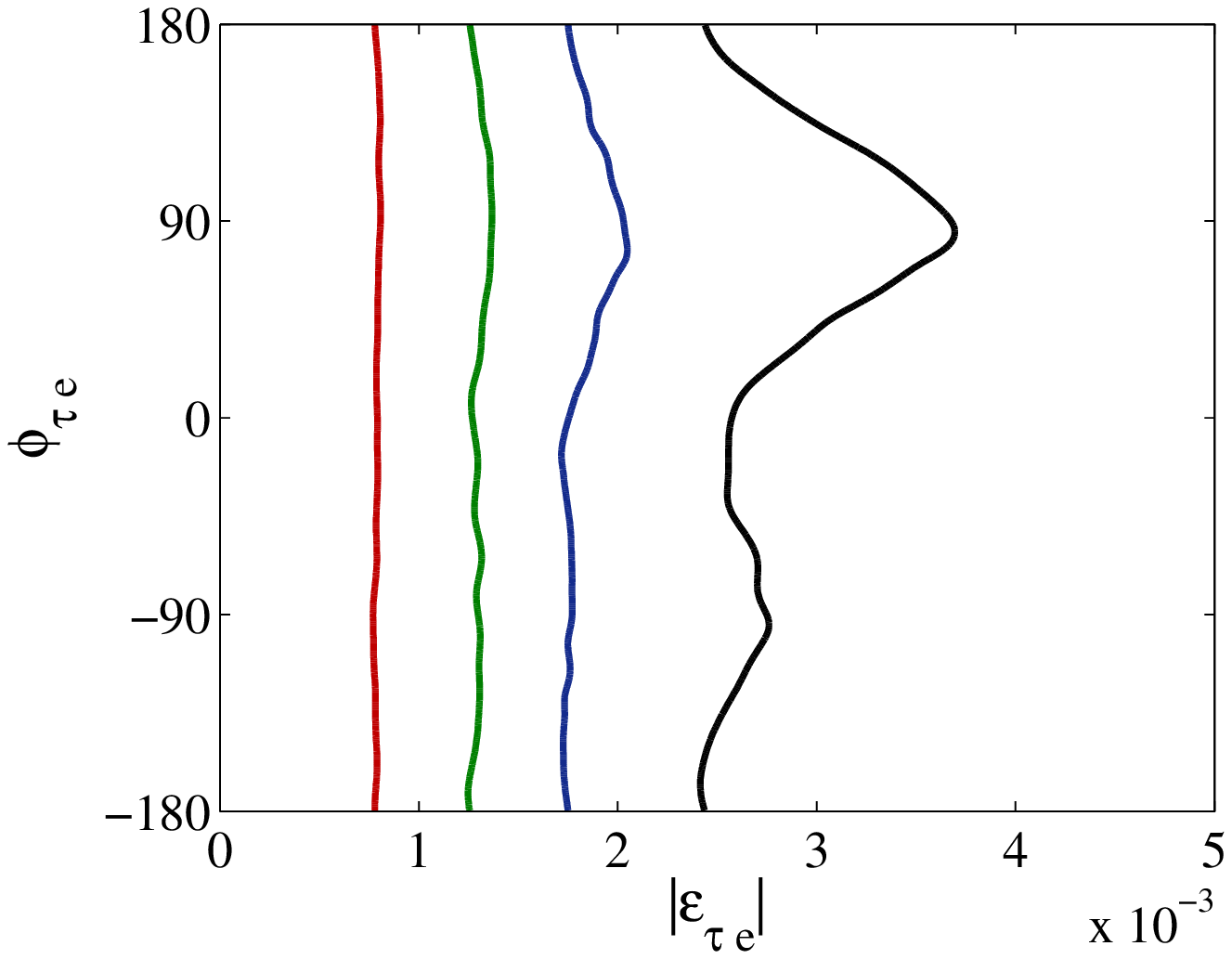}
 \end{center}
 \caption{\emph{The 90~\% confidence level sensitivity of the IDS Neutrino
   Factory to the unitarity violating parameters $\eps_{\tau\mu}$
   (left) and $\eps_{\tau e}$ (right). The different curves correspond
   to different sizes of the near $\tau$ detector, from left to right,
   10~kton, 1~kton, 100~ton, no near detector.}}
 \label{fig:tm-sens}
\end{figure}
In the left panel of \Fig~\ref{fig:tm-sens}, we show the sensitivity
to the $\eps_{\tau\mu}$ parameter for the four different sizes
considered for the near ECC.  The input values for all the
non-unitarity parameters and $\theta_{13}$ were set to zero to derive
these curves. We have checked that the results do not depend strongly
on this assumption.  The most remarkable feature of this figure is the
extreme sensitivity to the real part of $\eps_{\tau\mu}$ which is
present already without any near detector. This sensitivity mainly
originates from the matter effect on the disappearance channel, where
the leading non-unitarity correction to the oscillation probability is
given by
\begin{equation}
 \hat P_{\mu\mu} =  P_{\mu\mu}^{\rm SM}
 - 2\Real(\eps_{\mu\tau})AL\sin\left(\frac{\Delta m_{31}^2L}{2E}\right) + \mathcal O(\eps_{\mu \mu}),
\end{equation}
where $A = \sqrt{2}G_F n_e$, the terms we have omitted here can be found in
\App~\ref{sec:appendix}.  Notice that the discovery channel also
depends linearly on $\eps_{\tau \mu}$ and that the dependence is
\CP-violating. On the other hand, the mass and efficiency of the ECC
detector are much smaller compared to those of the MIND detectors for
the $\nu_\mu$ disappearance channel and therefore the sensitivity is
dominated by the latter.  As can be seen in the figure, a near $\tau$
detector will determine the modulus of $\eps_{\mu\tau}$ through the
zero-distance effect. This would translate into a vertical band in the
left panel of \Fig~\ref{fig:tm-sens} and thus the increase of the mass
of the near detector improves the measurement of the imaginary
part. However, given the linear dependence due to the matter effects
on propagation, the bound on the real part from the disappearance
channel remains stronger. We can also see that the bound on the
modulus does not require a very large near detector, the bound on the
imaginary part is essentially only improved by approximately 30~\% in
moving from a 1~kton to a 10~kton ECC detector.

\begin{figure}
 \begin{center}
  \includegraphics[width=0.49\textwidth]{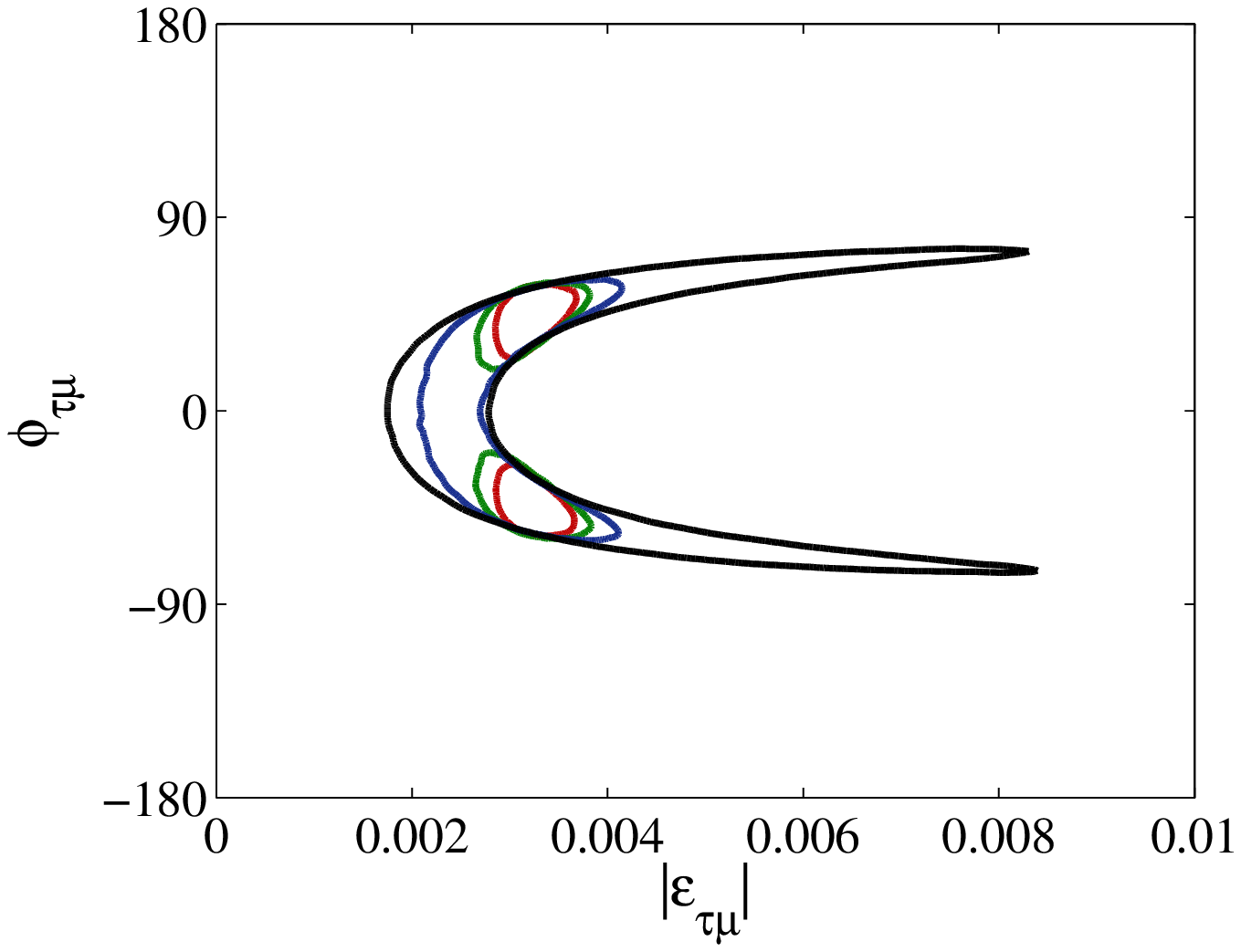}
  \includegraphics[width=0.49\textwidth]{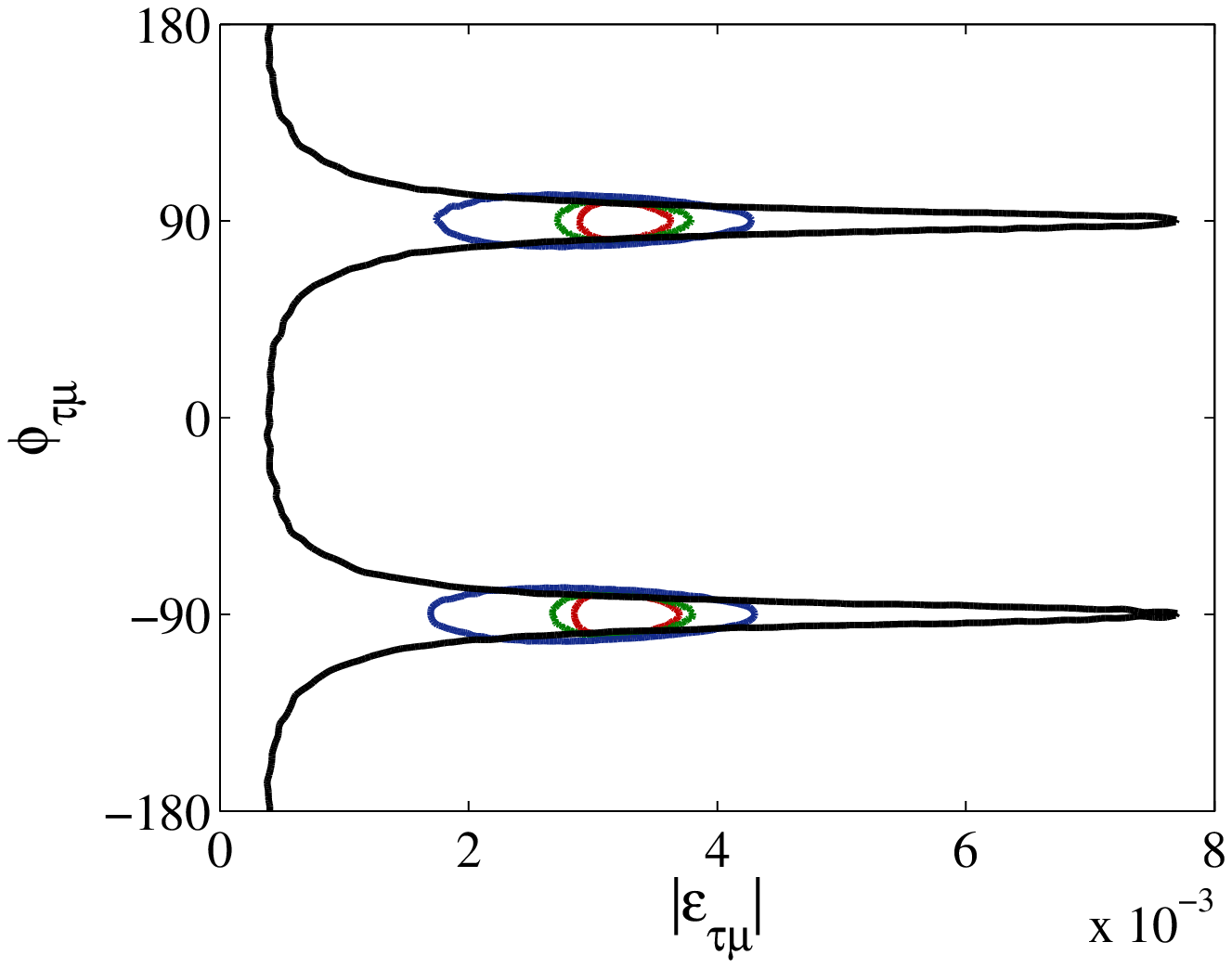}
 \end{center}
 \caption{\emph{The sensitivity of the IDS Neutrino Factory to the unitarity
   violating parameter $\eps_{\tau\mu}$, assuming that it takes the
   value $\eps_{\tau\mu} = 3.2\cdot 10^{-3} \exp(i\pi/4)$ (left) and
   $\eps_{\tau\mu} = - i\, 3.2\cdot 10^{-3}$ (right). The different curves
   correspond to different sizes of the near $\tau$ detector, from
   inner to outer curves, 10~kton, 1~kton, 100~ton, no near detector.}}
 \label{fig:tm-45}
\end{figure}

Another important question is how well the Neutrino Factory would be
able to measure the unitarity violating parameters if they are
non-zero. For this reason, in \Fig~\ref{fig:tm-45}, we show the
sensitivity to $\eps_{\tau\mu}$ assuming that $|\eps_{\tau\mu}| =
3.2\cdot 10^{-3}$ as well as $\phi_{\tau\mu} = 45^\circ$ and
$-90^\circ$, respectively, which is disfavoured at only $1\sigma$ by
current bounds. Thus, this gives a flavour of the best possible
situation for actually discovering unitarity violation and a new
source of \CP-violation.
Again, we can see that the sensitivity without the near detector is
only to the real part of $\eps_{\tau\mu}$. In this setting, there is a
degeneracy extending essentially as $|\eps_{\tau\mu}| \propto
1/\cos(\phi_{\tau\mu})$, along which the real part of $\eps_{\tau\mu}$
is constant and the imaginary part is changing. For the case with
purely imaginary $\eps_{\tau\mu}$ in the right panel of
\Fig~\ref{fig:tm-45}, it is also no surprise that the results without
the near detector are compatible with $\eps_{\tau\mu} = 0$. The
introduction of near detectors results in an effective measurement of
$|\eps_{\tau\mu}|$, \ie, a vertical band in the plot, which intersects
the far detector measurement giving rise to two degenerate solutions,
one for positive and one for negative imaginary part. Again, the
actual size of the near detector is not crucial and no significant
gain is seen beyond 1~kton.

These figures also show the strong complementarity between the near
and far detectors when it comes to measuring the phase of the
unitarity violating parameter, and thus also a non-standard source of
\CP-violation. Neither the near nor the far detectors alone can
establish a \CP-violating phase by themselves. However, combining the
two results excludes \CP-conservation at 90~\% confidence level.

Note that the slight widening of the allowed region when including the
near detector results from the use of Bayesian statistics. Since the
near detectors discard a large range of allowed values for
$\phi_{\tau\mu}$ when $|\eps_{\tau\mu}|$ is close to zero, a slightly
larger region in $\phi_{\tau\mu}$ close to the correct absolute value
of $\eps_{\tau\mu}$ is needed in order to include 90~\% of the
probability distribution.

\subsection{Sensitivity to $\boldsymbol{\eps_{\tau e}}$}

The right panel of \Fig~\ref{fig:tm-sens} shows the sensitivity to the
unitarity violation parameter $\eps_{\tau e}$ when the input values
for $\theta_{13}$ and all the unitarity violating parameters are set
to zero. Analogously to the sensitivity to $\eps_{\tau\mu}$, the setup
with only the far detectors is more sensitive to the real part of the
parameter, although the difference is not as pronounced. Furthermore,
as can be seen in the oscillation probabilities in
\App~\ref{sec:appendix}, the probabilities that depend on
$\eps_{\tau e}$ are only the golden, silver and discovery channels, where the
dependence is quadratic rather than linear, which translates into a
weaker bound. Thus, the inclusion of the near $\tau$ detector has a
major impact also on the bound which is placed on the real part of
$\eps_{\tau e}$. Indeed, for a 1~kton near $\tau$ detector, the
sensitivity is essentially flat as a function of $\phi_{\tau e}$ and
is dominated by the near detector.

\begin{figure}
 \begin{center}
  \includegraphics[width=0.49\textwidth]{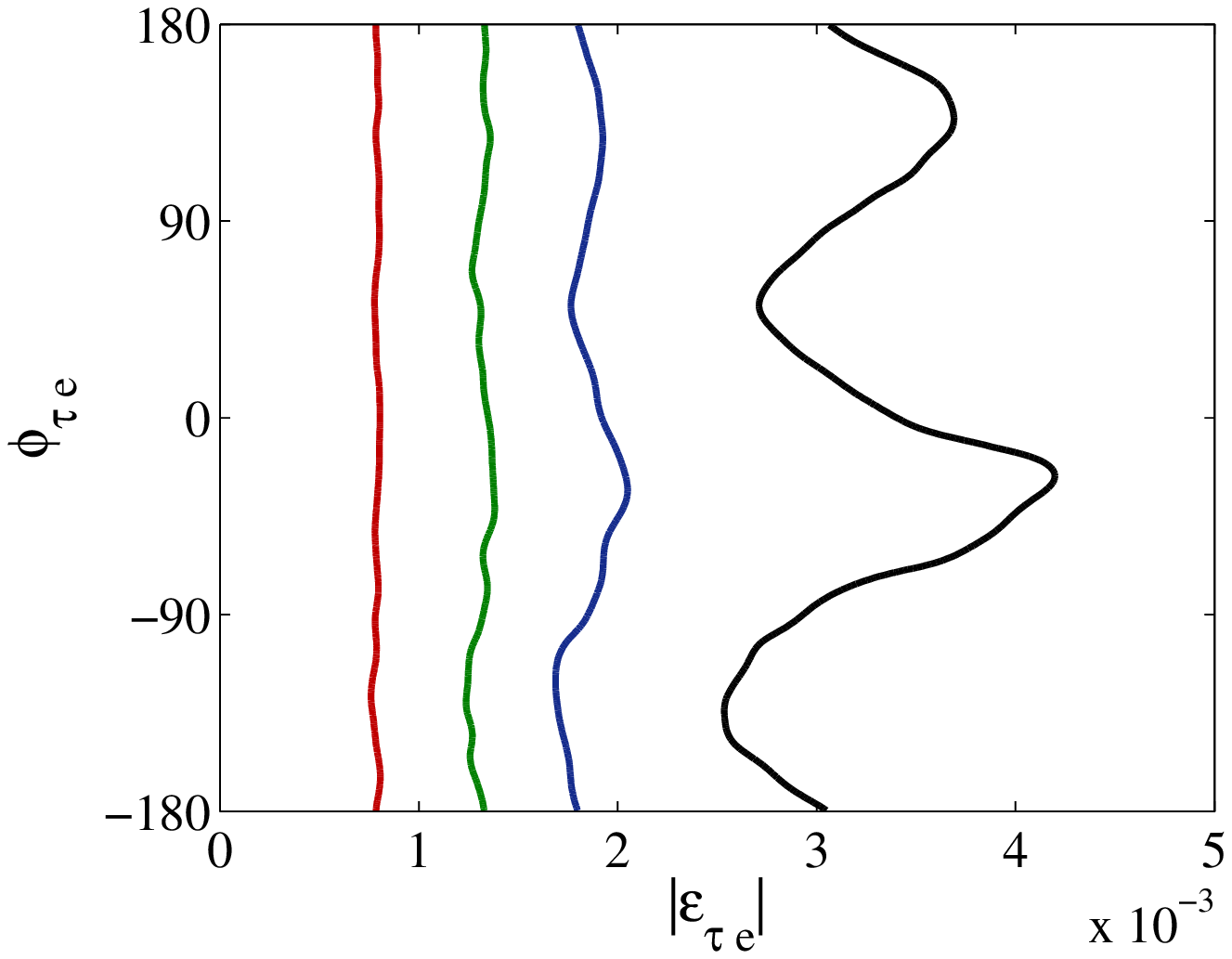}
  \includegraphics[width=0.49\textwidth]{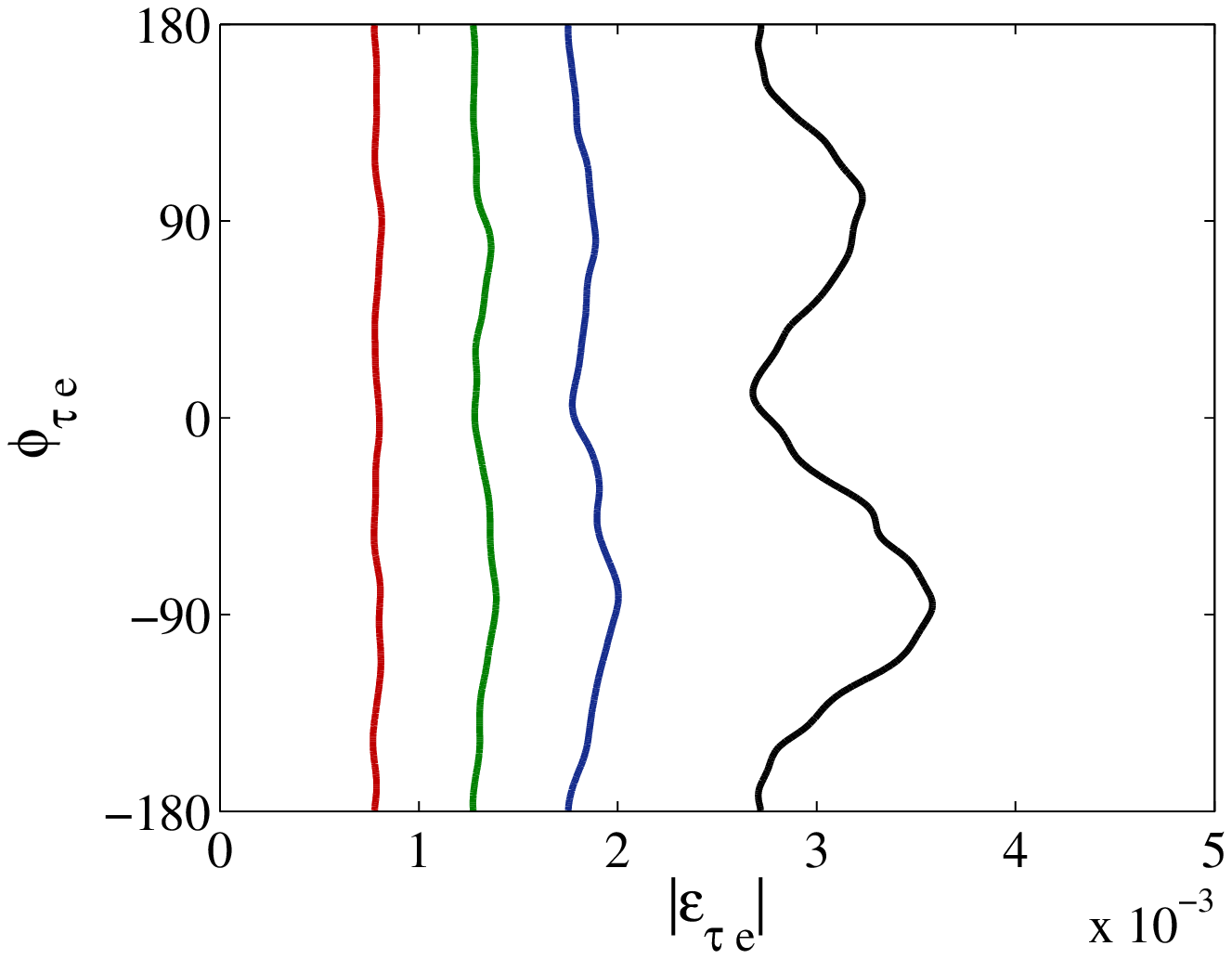}
 \end{center}
 \caption{\emph{The 90~\% confidence level sensitivity of the IDS
     Neutrino Factory to the unitarity violating parameter $\eps_{\tau
       e}$ with $\theta_{13}=5^\circ$ as well as $\delta=\pi/4$ (left) and
     $\delta=0$ (right).  The different curves correspond to different
     sizes of the near $\tau$ detector, from left to right, 10~kton,
     1~kton, 100~ton, no near detector.}}
 \label{fig:te-senst13}
\end{figure}

Again, the larger mass and efficiency of the MIND detector compared to
the ECC translates into the golden rather than the silver or the discovery channels
dominating the sensitivity to $\eps_{\tau e}$ from the far detectors
alone.  However, unlike the $\nu_\mu$ disappearance channel, the
golden channel is strongly dependent on the unknown parameters
$\theta_{13}$ and $\delta$ and the input values assumed for them will
influence the expected sensitivity to $\eps_{\tau e}$. Indeed, the
$\nu_e \to \nu_\mu$ probability in presence of non-unitarity is
modified to:
\begin{eqnarray}
\hat{P}_{e\mu}
&=&
P_{e\mu}^{\rm SM}+|\eps_{e\tau}|^2\sin^2\left(\frac{E_3L}{2}\right)
\nonumber\\
&&
+\Imag\left\lbrace \eps_{e\tau}\left[ \frac{1}{2}\frac{E_2}{A}\sin(2\theta_{12})
+\dfrac{E_3s_{13}e^{i\delta}}{A-E_3}\right] \right\rbrace\sin\left(\frac{AL}{2}\right)
\sin\left(\frac{E_3L}{2}\right)\sin\left(\frac{E_3-A}{2}L\right)
\nonumber\\
&&
+\Real\left\lbrace\eps_{e\tau}\left[ \frac{1}{\sqrt{2}}\frac{E_2}{A}\sin(2\theta_{12})
\sin\left(\frac{AL}{2}\right)\cos\left(\frac{E_3-A}{2}L\right)
\right.\right. \nonumber\\
&&
\left.\left.
\phantom{+\Real\{\eps_{e\tau}[}
-\dfrac{2\sqrt{2}E_3s_{13}e^{i\delta}}{A-E_3}\cos\left(\frac{AL}{2}\right)
\sin\left(\frac{E_3-A}{2}L\right)\right] \right\rbrace
\sin\left(\frac{E_3L}{2}\right) \nonumber\\
&&
+\mathcal{O}\left( \eps^3\right).
\end{eqnarray}
where $E_i = \Delta m_{i1}^2/(2E)$. It is then clear that the relative importance of the real and
imaginary parts of $\eps_{\tau e}$ in this probability strongly
depends on the actual values of $\theta_{13}$ and $\delta$. As an
example of this dependence, in \Fig~\ref{fig:te-senst13}, we again
show the sensitivity to $\eps_{\tau e}$, but for input values of
$\theta_{13}=5^\circ$ as well as for $\delta=\pi/4$ (left panel) and
$\delta=0$ (right panel).
Notice that while for $\delta=\pi/4$ the far MIND detectors are more
sensitive to the imaginary part of $\eps_{\tau e}$ the situation is
reversed for $\delta=0$.  However, the addition of the near $\tau$
detector for the silver channel dominates the bound and the curves
incorporating the near detectors forecast the same sensitivity
regardless of the true values of $\theta_{13}$ and $\delta$.

\begin{figure}
 \begin{center}
  \includegraphics[width=0.49\textwidth]{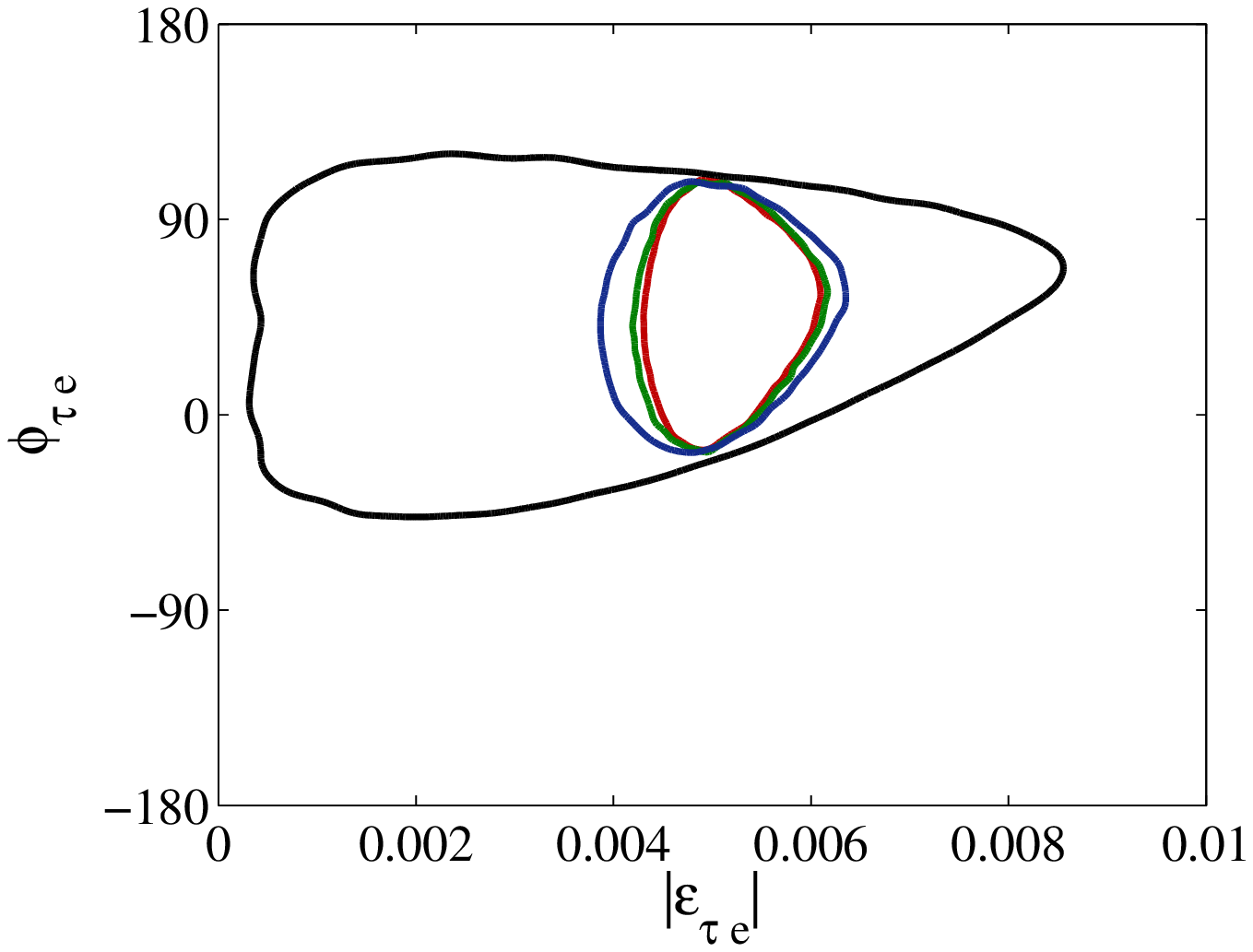}
  \includegraphics[width=0.49\textwidth]{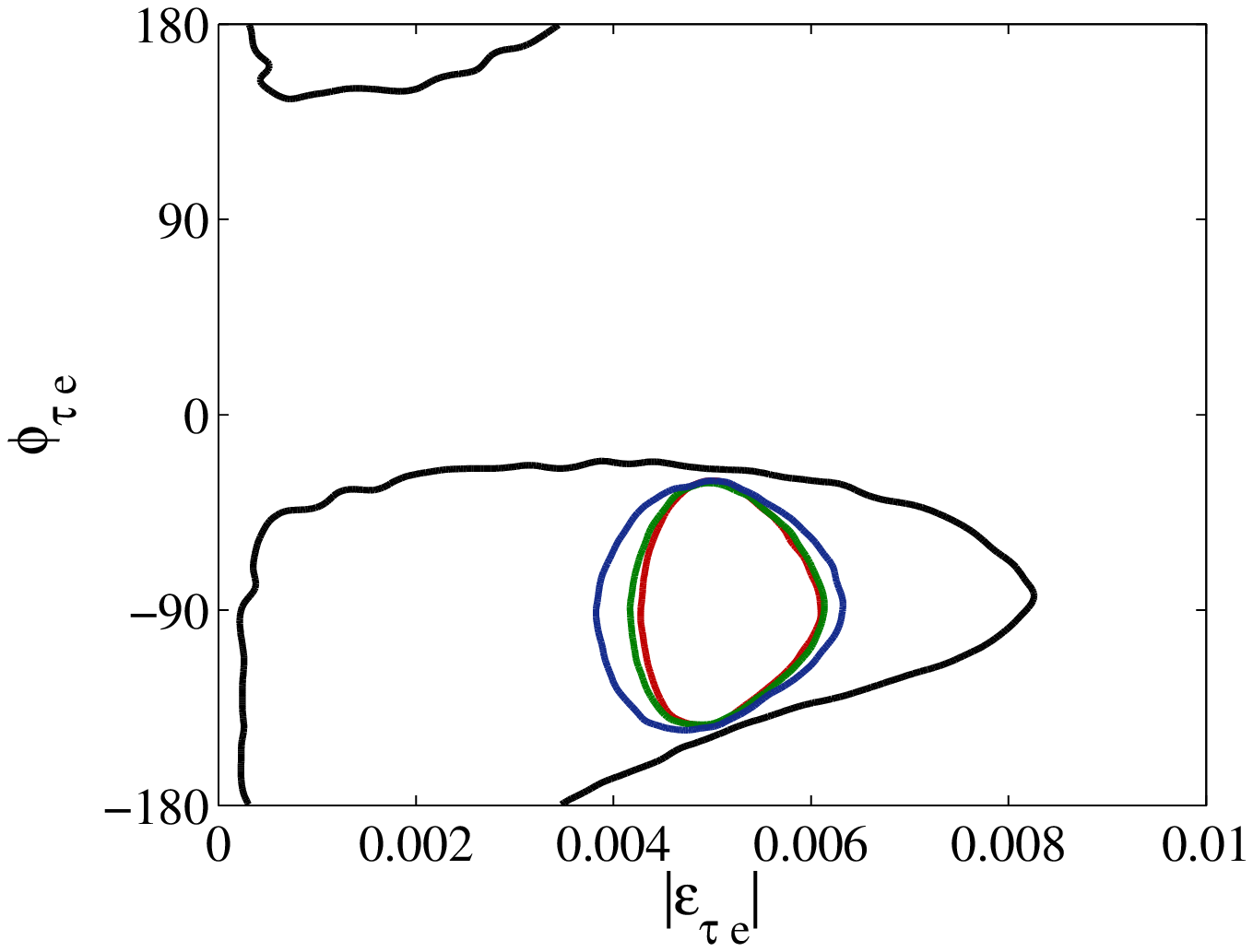}
 \end{center}
 \caption{\emph{The sensitivity of the IDS Neutrino Factory to the
     unitarity violating parameter $\eps_{\tau e}$, assuming that it
     takes the value $\eps_{\tau e} = 5.0\cdot 10^{-3} \exp(i\pi/4)$
     (left) and $\eps_{\tau\mu} = -i\, 5.0\cdot 10^{-3}$ (right). The
     different curves correspond to different sizes of the near $\tau$
     detector, from inner to outer curves, 10~kton, 1~kton, 100~ton,
     no near detector.}}
 \label{fig:te-45}
\end{figure}

In \Fig~\ref{fig:te-45}, we show the analogue of \Fig~\ref{fig:tm-45}
for $\epsilon_{\tau e}$. In this case, we assume $|\eps_{\tau e}| =
5.0\cdot 10^{-3}$ and $\phi_{\tau e} = 45^\circ$ and $-90^\circ$,
which again corresponds to the $1\sigma$ disfavoured region. For this
example, \CP-violation would not be discovered for the $\phi_{\tau e}
= 45^\circ$ case (left panel) at the $90$~\%~CL, but it would be
constrained around its true value already by the far detectors. In addition,
the inclusion of a near $\tau$ detector would again constrain the modulus and
therefore be complementary to the far detector result.
For the $\phi_{\tau e} = -90^\circ$ case (right panel), the complementarity of the near
and far detectors is able to exclude \CP-conservation at the 90~\%~CL.

\section{Summary and discussion}

We have considered the sensitivity of the IDS Neutrino Factory setup
to minimal unitarity violation (MUV) by using the Markov Chain Monte
Carlo methods implemented in MonteCUBES to explore the full parameter
space, consisting of the six standard neutrino oscillation parameters
and nine additional parameters describing the deviation from
unitarity. Our simulations were performed with several different near
ECC $\tau$ detector setups, ranging from no near detector to near
detector masses up to 10~kton.

Our results imply that the Neutrino Factory will be excellent for
probing some of the unitarity violating parameters. In particular,
a sensitivity of $\mathcal O(10^{-4})$ to the real part of the
unitarity violating parameter $\eps_{\tau\mu}$ is found. This is
mainly due to the matter effects in the $\nu_\mu$ disappearance
channel at the far detectors, for which the oscillation probability is
only linearly suppressed in $\Real(\eps_{\tau\mu})$. On the other
hand, we find that a near $\tau$ detector with a mass as small as
100~ton would dominate the sensitivity to $\eps_{\tau e}$, as well as
that to the imaginary part of $\eps_{\tau\mu}$, through the
measurement of the zero-distance effect, providing sensitivities down
to $\mathcal O(10^{-3})$.  For the other unitarity violating
parameters, we recover the priors of our simulation, which were set to
the current experimental bounds. The setup studied here will therefore
not improve our present knowledge of them.

Furthermore, we find no degeneracies neither among the different
unitarity violating parameters, nor between the unitarity violating
parameters and the small standard neutrino oscillation parameters,
such as $\theta_{13}$. This means that the sensitivities to the
standard oscillation parameters are robust even in presence of
unitarity violation.

Regarding the prospects of an actual detection of unitarity violation,
and especially \CP-violation stemming from non-unitary mixing, we find
that the near and far detectors play a very complementary role. In the
case of $\eps_{\tau\mu}$, the far detectors are only sensitive to the
real part of the unitarity violating parameter while the near detector
can measure its modulus, neither is sensitive to unitarity violating
\CP-violation by themselves. However, it can be effectively probed by
considering the combination of the two, as illustrated in
\Fig~\ref{fig:tm-45}.

We would like to stress that, while the sensitivity to unitarity
violation at a Neutrino Factory has been studied before
\cite{Campanelli:2002cc,FernandezMartinez:2007ms,Altarelli:2008yr,Goswami:2008mi,Malinsky:2009gw},
the sensitivity to the real part of $\eps_{\tau\mu}$ due to matter
effects has not been discussed (however, a similar term in the
$\nu_\mu$ disappearance channel is present in and has been studied for
the case of oscillations into sterile
neutrinos~\cite{Donini:2008wz}). Furthermore, these studies have not
systematically scanned the parameter space while keeping all
parameters free within their prior values. Thus, the observation that
there are no extended degeneracies, neither between the standard and
unitarity violating parameters, nor among the unitarity violating
parameters themselves, is also new.

We conclude that a Neutrino Factory would provide powerful tool for
probing unitarity violation in the leptonic mixing matrix. For the
parameters to which it is most sensitive, the sensitivity is an order
of magnitude better than the current experimental bounds. Finally, the
interplay between the near and far detectors would allow to test new
sources of \CP-violation in the lepton sector.

\begin{acknowledgments}
We would like to thank A.~Donini, W.~Wang, and Z.-z.~Xing
for useful discussions.
This work was supported by the Swedish Research Council
(Vetenskapsr\aa{}det), contract no.\ 623-2007-8066 [M.B.]. S.A., M.B.\ and E.F.M.\
acknowledge support by the DFG cluster of excellence ``Origin
and Structure of the Universe''. J.L.P.\ acknowledges financial support by the Ministry of Science and Innovation of Spain (MICIINN) through an FPU grant, ref.~AP2005-1185. S.A., E.F.M.\ and J.L.P.\ also acknowledge support
from the European Community under the European Commission Framework Programme
7 Design Study: EUROnu, Project Number 212372.
\end{acknowledgments}

\appendix

\section{OSCILLATION PROBABILITIES IN THE PRESENCE OF UNITARITY VIOLATION}
\label{sec:appendix}
\providecommand{\ket}[1]{\left|#1\right>}

In this Appendix, we derive the probabilities $P_{\alpha\beta}$ in
matter assuming constant density. In order to perform the calculation,
we will use the Kimura--Takamura--Yokomura (KTY)
formalism~\cite{Kimura:2002wd,Yasuda:2007jp}, which has already been
applied to the MUV scheme in the Appendix of
\Ref~\cite{FernandezMartinez:2007ms}.
Since the constraint on $\eps_{e\mu}$ is strong enough to safely
neglect $\eps_{e\mu}$ in the oscillation probabilities, we will not
consider it below. However, it has been considered in the numerical
analysis presented in the main part of this paper. The effective
flavour eigenstates are given by: 
\begin{equation}
\ket{\nu_\alpha} =
\dfrac{(1+\eps^*)_{\alpha \beta}U_{\beta i}^*}{\left[
    1+2\eps_{\alpha\alpha}+(\eps^2)_{\alpha\alpha}\right]^{1/2}}\,\ket{\nu_i}
\equiv \dfrac{(1+\eps^*)_{\alpha \beta}}{\left[
    1+2\eps_{\alpha\alpha}+(\eps^2)_{\alpha\alpha}\right]^{1/2}
}\,\ket{\nu^{SM}_\beta} .
\label{estadosjuntos}
\end{equation}
The parameters that appear linearly in the normalisation factors are
$\eps_{ee}$, $\eps_{\mu\mu}$, and $\eps_{\tau\tau}$, which are already
better constrained by other considerations than the sensitivities we
find for a Neutrino Factory. Thus, the determination of the fluxes and
cross-sections by the near detectors only suffer from a minor
additional theoretical uncertainty.  We will present the oscillation
probabilities $\hat{P}(\nu_\alpha\rightarrow\nu_\beta) = \hat
P_{\alpha\beta}$ without taking the normalisation factors into
account.
Notice that this will not be at all relevant for the golden and silver
channels, since the probabilities are already order $\eps^2$ before
taking the normalization factors into account. Thus, the corrections
would be at most $\mathcal O(\eps^3)$.

The oscillation probability $\hat P_{\alpha\beta}$, expressed as a
function of the KTY parameters, is~\cite{FernandezMartinez:2007ms}:
\begin{eqnarray}
\hat{P}_{\alpha\beta}
&=&
|(NN^{\dagger})_{\alpha\beta}|^2
-4\sum_{j<k}\mbox{\rm Re}(\tilde{X}^{\alpha\beta}_j
\tilde{X}^{\alpha\beta\ast}_k)
\sin^2\left(\frac{\Delta \tilde{E}_{jk}L}{2}\right)\nonumber\\
&&
+2\sum_{j<k}\mbox{\rm Im}(\tilde{X}^{\alpha\beta}_j
\tilde{X}^{\alpha\beta\ast}_k)
\sin(\Delta \tilde{E}_{jk}L),
\label{modifiedprob}
\end{eqnarray}
where $\Delta \tilde{E}_{jk}\equiv \tilde{E}_{j}-\tilde{E}_{k}$ and
$\tilde{X}^{\alpha\beta}_j\equiv (N^{*}W)_{\alpha j}(N^{*}W)^{*}_{\beta
  j}$~($j=1,2,3)$. Here, $\tilde E_i$ are the effective eigenvalues in
matter and $W_{ij}$ is the unitary matrix which diagonalizes the evolution equation for the mass eigenstates:

\bea
i\frac{d}{dt}\ket{\nu_i}&=&
\left[ {\mbox{\rm diag}}(E_1,E_2,E_3)+N^\dagger\,\sqrt{2}G_F{\mbox{\rm diag}}
(n_e-n_n/2,-n_n/2,-n_n/2)\, N\right]_{ji} \ket{\nu_j}
\nonumber\\
&\equiv&{\cal H}_{ji}\ket{\nu_j}
\eea
where $E_i = \Delta m_{i1}^2/(2E)$.  Assuming that the electron and
neutron number densities are equal\footnote{This is a very good
  approximation in the case of neutrino oscillations in the Earth.}
(\ie, $n_e = n_n$), ${\cal H}$ can be expressed as
\begin{equation}
{\cal H}=
\diag(E_1,E_2,E_3)+N^\dagger \diag\left(\frac{A}{2},-\frac{A}{2},-\frac{A}{2}\right) N,
\end{equation}
where $A=\sqrt{2}G_F n_e$. Finally, according to the KTY formalism
applied to the MUV scheme (again, see
\Ref~\cite{FernandezMartinez:2007ms}), $\tilde{X}^{\alpha\beta}_j$ can
be expressed as
\begin{equation}
 \tilde{X}^{\alpha\beta}_j\equiv\sum_{l}\left(V^{-1}\right)_{jl}Y^{\alpha\beta}_l
  =\sum_{l}\left(V^{-1}\right)_{jl}\,\left[ N \,{\cal H}^{l-1}\,N^\dagger \right]_{\beta\alpha}, 
 \label{X's}
\end{equation}
where
\medskip
\begin{eqnarray}
 V^{-1} =
 \begin{pmatrix}
  (\Delta\tilde{E}_{21}\Delta\tilde{E}_{31})^{-1}\,(\tilde{E}_{2}\tilde{E}_{3},-\tilde{E}_{2}-\tilde{E}_{3},1) \\  
 -(\Delta\tilde{E}_{21}\Delta\tilde{E}_{32})^{-1}\,(\tilde{E}_{3}\tilde{E}_{1},-\tilde{E}_{3}-\tilde{E}_{1},1) \\ 
  (\Delta\tilde{E}_{31}\Delta\tilde{E}_{32})^{-1}\,(\tilde{E}_{2}\tilde{E}_{1},-\tilde{E}_{2}-\tilde{E}_{1},1)
 \end{pmatrix} \,.
 \label{limits}
\end{eqnarray}
Once the effective eigenvalues in matter are known, it is
straightforward to obtain the expressions for the neutrino oscillation
probabilities. However, in order to obtain reasonably simple
expressions, it is necessary to expand them in small parameters. Here,
we present the oscillation probabilities to second order in the
parameters listed in \Tab~\ref{tab:expansionparameters}.
\begin{table}
 \begin{center}
  \begin{tabular}{|c|c|}
   \hline
   SM expansion parameters ($\eta$) & MUV expansion parameters \\
   \hline
   $\theta_{13}$, $\Delta m_{21}^2/\Delta m_{31}^2$, $\delta\theta_{23} = \theta_{23} - \pi/4$
   &
   $\eps_{\alpha\beta}$ \\
   \hline
  \end{tabular}
 \end{center}
 \caption{\emph{The small expansion parameters used in our neutrino
     oscillation probabilities. We will refer to the set of SM
     expansion parameters as $\eta$. The full set of expansion
     parameters will be referred to as $\eps$, while only the set of
     MUV expansion parameters will be denoted by
     $\eps_{\alpha\beta}$.}}
 \label{tab:expansionparameters}
\end{table}

To second order in $\eps$, we can find the eigenvalues by using
perturbation theory. We find that
\begin{eqnarray}
\tilde{E}_1
 &=&
  A\left[ 1+\frac{E_2}{A}s_{12}^2+\frac{1}{4}\frac{E_2^2}{A^2}\sin^2(2\theta_{12})
   +\frac{E_3s_{13}^2}{A-E_3}+\eps_{ee}+\frac{\eps_{ee}^2}{2}-\frac{|\eps_{e\tau}|^2}{2}\right]+\mathcal{O}(\eps^3)\,,
\\
\tilde{E}_2
 &=&
  A\left\{\frac{E_2}{A}c_{12}^2-\frac{E_2^2}{4A^2}\sin^2(2\theta_{12})
  +\Real\left(\eps_{\mu\tau}\right)\left[1+\frac{1}{2}(\eps_{\mu\mu}+\eps_{\tau\tau})\right]
  -\frac{1}{2}(\eps_{\mu\mu}+\eps_{\tau\tau})\right. 
  \nonumber\\
 &&
 \phantom{A}\left.-\frac{1}{4}(\eps_{\mu\mu}^2+\eps_{\tau\tau}^2)
  -\frac{|\eps_{\mu\tau}|^2}{2}+\frac{|\eps_{e\tau}|^2}{4}-\delta\theta_{23}
  \left[ \eps_{\tau\tau}-\eps_{\mu\mu}+\frac{1}{2}(\eps_{\tau\tau}^2-\eps_{\mu\mu}^2)-|\eps_{e\tau}|^2/2\right] \right.
  \nonumber\\
 &&
 \phantom{A}
  -\frac{A}{E_3}\Real(\eps_{\mu\tau})^2-\left.\frac{A}{4E_3}(\eps_{\tau\tau}-\eps_{\mu\mu})^2\right\}
  +\mathcal{O}(\eps^3)\,,
 \\
\tilde{E}_3
 &=&
  A\left\{\frac{E_3}{A}-\frac{E_3 s_{13}^2}{A-E_3}-\Real\left(\eps_{\mu\tau}\right)
  \left[1+\frac{1}{2}(\eps_{\mu\mu}+\eps_{\tau\tau})\right]+\delta\theta_{23}(\eps_{\tau\tau}-\eps_{\mu\mu})\right.
  \nonumber\\
 &&
 \phantom{A}
  -\left.\frac{1}{2}(\eps_{\mu\mu}+\eps_{\tau\tau})-\frac{1}{4}(\eps_{\tau\tau}^2+\eps_{\mu\mu}^2)
  -\frac{|\eps_{\mu\tau}|^2}{2}+\frac{|\eps_{e\tau}|^2}{4}\right\}+\mathcal{O}(\eps^3)\,.
\label{eigenvalues}
\end{eqnarray} 
Notice that, for $\eps_{\alpha\beta} \rightarrow 0$, we recover the SM
results as expected. These results allow us to obtain $V^{-1}$ at
second order. Thus, we only need to compute $Y^{\alpha\beta}_j$ at the
same order, the computation is straightforward but tedious (see
\eq~(\ref{X's})). For brevity, we do not present the results for
$V^{-1}$ and $Y^{\alpha\beta}_j$ here. However, we would like to
comment that, for the golden and silver channels, it is enough to
compute these quantities to first order, since
$\tilde{X}^{\alpha\beta}_j$ is already of first order in $\eta$. This
is not true in the case of the $\nu_\mu$-$\nu_\tau$ sector, where
$\tilde{X}^{\mu\mu}_2|_{\eps=0}=\tilde{X}^{\mu\mu}_3|_{\eps=0}=-\tilde{X}^{\tau\mu}_2|_{\eps=0}=\tilde{X}^{\tau\mu}_3|_{\eps=0}=1/2$. The
advantage of this sector, from the point of view of discovering new
physics, is that the effects of the new physics can appear in the
probability at first order as an interference term between the SM and
the new physics without additional suppression by $\eta$. For this
reason, we keep only the interference between the $\mathcal{O}\left(
\eps_{\alpha\beta}\right)$ terms and the $\mathcal{O}\left(
\eta\right)$ ones at second order\footnote{It could also be justified
  to neglect the $\mathcal{O}\left(\eps_{\alpha\beta}\dfrac{\Delta
    m_{21}^2}{\Delta m_{31}^2}\right)$ terms, since the maximal
  allowed value of $\dfrac{\Delta m_{21}^2}{\Delta m_{31}^2}$ is at
  least one order of magnitude smaller than the maximal allowed values
  of $s_{13}$ and $\delta\theta_{23}$. However, we keep also these
  terms for completeness.} in that sector.

In the end, we obtain the following expanded oscillation probabilities
at the orders mentioned above:
\begin{eqnarray}
%
%
\hat{P}_{\mu\mu}
 &=&
  P_{\mu\mu}^{\rm SM}+4\eps_{\mu\mu}+4\eps_{\mu\mu}^2
  \nonumber\\
 &&
  + 4\left\lbrace -\eps_{\mu\mu}+2\Real(\eps_{\mu\tau})\delta\theta_{23}
  -2\delta\theta_{23}(\eps_{\mu\mu}-\eps_{\tau\tau})\frac{A}{E_3}\right\rbrace
  \sin^2\left(\dfrac{E_3L}{2}\right)  
  \nonumber\\
 &&
  -\left[ 2\Real(\eps_{\mu\tau})-\delta\theta_{23}(\eps_{\mu\mu}-\eps_{\tau\tau})\right]  AL\sin(E_3L)
  +\mathcal{O}\left( \eps_{\alpha\beta}^2\right)\, ,
  \\
  \nonumber\\
%
%
\hat{P}_{\mu\tau}
 &=&
  P_{\mu\tau}^{\rm SM}+4|\eps_{\mu\tau}|^2
  \nonumber\\
 &&
  +\left[ 2\Real(\eps_{\mu\mu}+\eps_{\tau\tau})+8\delta\theta_{23}(\eps_{\mu\mu}-\eps_{\tau\tau})
  \frac{A}{E_3}\right]\sin^2\left(\dfrac{E_3L}{2}\right)
  \nonumber\\
 &&
  +\left[ -2\Imag(\eps_{\mu\tau})-\delta\theta_{23}(\eps_{\mu\mu}-\eps_{\tau\tau})AL\right] \sin(E_3L)
  \nonumber\\
 &&
  -\sqrt{2}\Imag\left\lbrace\eps_{e\tau}\left[\frac{E_2}{A}\sin(2\theta_{12})
  +\dfrac{2E_3s_{13}e^{i\delta}}{A-E_3}\right] \right\rbrace \sin\left(\frac{AL}{2}\right)
  \sin\left(\frac{E_3L}{2}\right)\sin\left(\frac{E_3-A}{2}L\right)
  \nonumber\\
 &&
  +\sqrt{2}\Real\left\lbrace\eps_{e\tau}\left[\frac{E_2}{A}\sin(2\theta_{12})
  \sin\left(\frac{AL}{2}\right)\cos\left(\frac{E_3-A}{2}L\right) \right.\right.
  \nonumber\\
 &&
 \phantom{+\sqrt 2 \Imag \{\eps_{e\tau}[}
  -\left.\left.\dfrac{2E_3s_{13}e^{i\delta}}{A-E_3}\cos\left(\frac{AL}{2}\right)
  \sin\left(\frac{E_3-A}{2}L\right)\right]  \right\rbrace\sin\left(\frac{E_3L}{2}\right) 
  \nonumber\\
 &&
  +\mathcal{O}\left( \eps_{\alpha\beta}^2\right)\, ,
  \\
  \nonumber\\
%
%
\hat{P}_{e\mu}
 &=&
  P_{e\mu}^{\rm SM}+|\eps_{e\tau}|^2\sin^2\left(\frac{E_3L}{2}\right)
  \nonumber\\
 &&
  +\Imag\left\lbrace \eps_{e\tau}\left[ \frac{1}{2}\frac{E_2}{A}\sin(2\theta_{12})
  +\dfrac{E_3s_{13}e^{i\delta}}{A-E_3}\right] \right\rbrace\sin\left(\frac{AL}{2}\right)
  \sin\left(\frac{E_3L}{2}\right)\sin\left(\frac{E_3-A}{2}L\right)
  \nonumber\\
 &&
  +\Real\left\lbrace\eps_{e\tau}\left[ \frac{1}{\sqrt{2}}\frac{E_2}{A}\sin(2\theta_{12})
  \sin\left(\frac{AL}{2}\right)\cos\left(\frac{E_3-A}{2}L\right)\right.\right.
  \nonumber\\
 &&
 \left.\left.
 \phantom{+\Real\{\eps_{e\tau}[}
  -\dfrac{2\sqrt{2}E_3s_{13}e^{i\delta}}{A-E_3}\cos\left(\frac{AL}{2}\right)
  \sin\left(\frac{E_3-A}{2}L\right)\right] \right\rbrace
  \sin\left(\frac{E_3L}{2}\right)
  \nonumber\\
 &&
  +\mathcal{O}\left( \eps^3\right)\, ,
  \\
  \nonumber\\
%
%
\hat{P}_{e\tau}
 &=&
  P_{e\tau}^{\rm SM}
  + 4|\eps_{e\tau}|^2-2\left[ |\eps_{e\tau}|^2
  -\dfrac{\sqrt{2}E_3s_{13}}{A-E_3}\Real(\eps_{e\tau}e^{i\delta})\right]
  \sin^2\left(\frac{E_3-A}{2}L\right)
  \nonumber\\
 &&
  -2\left[ |\eps_{e\tau}|^2-\frac{1}{\sqrt{2}}\frac{E_2}{A}\sin(2\theta_{12})
  \Real(\eps_{e\tau})\right]\sin^2\left(\frac{AL}{2}\right)
  \nonumber\\
 &&
  -\Imag\left\lbrace \eps_{e\tau}^{*}\left[ \frac{1}{\sqrt{2}}\frac{E_2}{A}
  \sin(2\theta_{12})\sin(AL)-\dfrac{\sqrt{2}E_3s_{13}e^{-i\delta}}{A-E_3}\sin(\{E_3-A\}L)\right] \right\rbrace
  \nonumber\\
 &&
  -2\sqrt{2}\Real\left\lbrace \eps_{e\tau}\left[ \frac{1}{2}\frac{E_2}{A}\sin(2\theta_{12})
  -\dfrac{E_3s_{13}e^{i\delta}}{A-E_3}\right] \right\rbrace\sin\left(\frac{AL}{2}\right)
  \cos\left(\frac{E_3L}{2}\right)\sin\left(\frac{E_3-A}{2}L\right)
  \nonumber\\
 &&
  +\Imag\left\lbrace\eps_{e\tau}\left[
  \sqrt{2}\frac{E_2}{A}\sin(2\theta_{12})\sin\left(\frac{AL}{2}\right)
  \cos\left(\frac{E_3-A}{2}L\right) \right.\right.
  \nonumber\\
 &&
 \left.\left.
 \phantom{-\Imag\{\eps_{e\tau}[}
  +\dfrac{2\sqrt{2}E_3s_{13}e^{i\delta}}{A-E_3}\cos\left(\frac{AL}{2}\right)
  \sin\left(\frac{E_3-A}{2}L\right) \right] \right\rbrace 
  \cos\left(\frac{E_3L}{2}\right)
  \nonumber\\
 &&
  +\mathcal{O}\left(\eps^3\right)\, .
\end{eqnarray}
Notice that we do not neglect the zero-distance effect in the
$\nu_\mu$-$\nu_\tau$ sector. Although this is not within the order of
the expansion, we keep it as it plays an important role in the
analysis of the neutrino flavour transitions at near detectors.


\begin{thebibliography}{10}

\bibitem{Geer:1997iz}
S.~Geer,
\newblock Phys. Rev. {\bf D57}, 6989 (1998), [hep-ph/9712290].

\bibitem{DeRujula:1998hd}
A.~De~Rujula, M.~B. Gavela and P.~Hernandez,
\newblock Nucl. Phys. {\bf B547}, 21 (1999), [hep-ph/9811390].

\bibitem{Langacker:1988ur}
P.~Langacker and D.~London,
\newblock Phys. Rev. {\bf D38}, 886 (1988).

\bibitem{Antusch:2006vwa}
S.~Antusch, C.~Biggio, E.~Fernandez-Martinez, M.~B. Gavela and J.~Lopez-Pavon,
\newblock JHEP {\bf 10}, 084 (2006), [hep-ph/0607020].

\bibitem{Abada:2007ux}
A.~Abada, C.~Biggio, F.~Bonnet, M.~B. Gavela and T.~Hambye,
\newblock JHEP {\bf 12}, 061 (2007), [0707.4058].

\bibitem{FernandezMartinez:2007ms}
E.~Fernandez-Martinez, M.~B. Gavela, J.~Lopez-Pavon and O.~Yasuda,
\newblock Phys. Lett. {\bf B649}, 427 (2007), [hep-ph/0703098].

\bibitem{Nardi:1994iv}
E.~Nardi, E.~Roulet and D.~Tommasini,
\newblock Phys. Lett. {\bf B327}, 319 (1994), [hep-ph/9402224].

\bibitem{Tommasini:1995ii}
D.~Tommasini, G.~Barenboim, J.~Bernabeu and C.~Jarlskog,
\newblock Nucl. Phys. {\bf B444}, 451 (1995), [hep-ph/9503228].

\bibitem{Antusch:2008tz}
S.~Antusch, J.~P. Baumann and E.~Fernandez-Martinez,
\newblock Nucl. Phys. {\bf B810}, 369 (2009), [0807.1003].

\bibitem{Bekman:2002zk}
B.~Bekman, J.~Gluza, J.~Holeczek, J.~Syska and M.~Zralek,
\newblock Phys. Rev. {\bf D66}, 093004 (2002), [hep-ph/0207015].

\bibitem{Holeczek:2007kk}
J.~Holeczek, J.~Kisiel, J.~Syska and M.~Zralek,
\newblock Eur. Phys. J. {\bf C52}, 905 (2007), [0706.1442].

\bibitem{Xing:2007zj}
Z.-z. Xing,
\newblock Phys. Lett. {\bf B660}, 515 (2008), [0709.2220].

\bibitem{Xing:2009vb}
Z.-z. Xing,
\newblock [0902.2469].

\bibitem{Broncano:2002rw}
A.~Broncano, M.~B. Gavela and E.~E. Jenkins,
\newblock Phys. Lett. {\bf B552}, 177 (2003), [hep-ph/0210271].

\bibitem{Bandyopadhyay:2007kx}
ISS Physics Working Group, A.~Bandyopadhyay {\em et~al.},
\newblock [0710.4947].

\bibitem{IDS}
{IDS} homepage,
\newblock https://www.ids-nf.org/wiki/FrontPage.

\bibitem{Abe:2007bi}
{ISS Detector Working Group}, T.~Abe {\em et~al.},
\newblock [0712.4129].

\bibitem{Donini:2002rm}
A.~Donini, D.~Meloni and P.~Migliozzi,
\newblock Nucl. Phys. {\bf B646}, 321 (2002), [hep-ph/0206034].

\bibitem{Autiero:2003fu}
D.~Autiero {\em et~al.},
\newblock Eur. Phys. J. {\bf C33}, 243 (2004), [hep-ph/0305185].

\bibitem{BurguetCastell:2001ez}
J.~Burguet-Castell, M.~B. Gavela, J.~J. Gomez-Cadenas, P.~Hernandez and
  O.~Mena,
\newblock Nucl. Phys. {\bf B608}, 301 (2001), [hep-ph/0103258].

\bibitem{Huber:2003ak}
P.~Huber and W.~Winter,
\newblock Phys. Rev. {\bf D68}, 037301 (2003), [hep-ph/0301257].

\bibitem{Tang:2009na}
J.~Tang and W.~Winter,
\newblock [0903.3039].

\bibitem{Cervera:2000kp}
A.~Cervera {\em et~al.},
\newblock Nucl. Phys. {\bf B579}, 17 (2000), [hep-ph/0002108].

\bibitem{Donini:2008wz}
A.~Donini, K.-i. Fuki, J.~Lopez-Pavon, D.~Meloni and O.~Yasuda,
\newblock [0812.3703].

\bibitem{mcubes}
M.~Blennow and E.~Fernandez-Martinez,
\newblock [0903.3985].

\bibitem{mcubeshome}
{MonteCUBES} homepage,
\newblock http://wwwth.mppmu.mpg.de/members/blennow/montecubes/.

\bibitem{Huber:2004ka}
P.~Huber, M.~Lindner and W.~Winter,
\newblock Comput. Phys. Commun. {\bf 167}, 195 (2005), [hep-ph/0407333].

\bibitem{Huber:2007ji}
P.~Huber, J.~Kopp, M.~Lindner, M.~Rolinec and W.~Winter,
\newblock Comput. Phys. Commun. {\bf 177}, 432 (2007), [hep-ph/0701187].

\bibitem{Gelman:1992}
A.~Gelman and D.~Rubin,
\newblock Statistical Science {\bf 7}, 457 (1992).

\bibitem{Maltoni:2004ei}
M.~Maltoni, T.~Schwetz, M.~A. Tortola and J.~W.~F. Valle,
\newblock New J. Phys. {\bf 6}, 122 (2004), [hep-ph/0405172].

\bibitem{GonzalezGarcia:2007ib}
M.~C. Gonzalez-Garcia and M.~Maltoni,
\newblock Phys. Rept. {\bf 460}, 1 (2008), [0704.1800].

\bibitem{Campanelli:2002cc}
M.~Campanelli and A.~Romanino,
\newblock Phys. Rev. {\bf D66}, 113001 (2002), [hep-ph/0207350].

\bibitem{Altarelli:2008yr}
G.~Altarelli and D.~Meloni,
\newblock Nucl. Phys. {\bf B809}, 158 (2009), [0809.1041].

\bibitem{Goswami:2008mi}
S.~Goswami and T.~Ota,
\newblock Phys. Rev. {\bf D78}, 033012 (2008), [0802.1434].

\bibitem{Malinsky:2009gw}
M.~Malinsky, T.~Ohlsson and H.~Zhang,
\newblock [0903.1961].

\bibitem{Kimura:2002wd}
K.~Kimura, A.~Takamura and H.~Yokomakura,
\newblock Phys. Rev. {\bf D66}, 073005 (2002), [hep-ph/0205295].

\bibitem{Yasuda:2007jp}
O.~Yasuda,
\newblock [0704.1531].

\end{thebibliography}

\end{document}